\documentclass[twocolumn,showpacs,prb]{revtex4}
\usepackage{amsmath}
\usepackage{graphicx}

\begin{document}
\title{Electric field gradient induced effects in GaAs/AlGaAs modulation-doped
structures and high frequency sensing}

\author{D.~Seliuta, A.~Juozapavi\v{c}ius, V.~Gru\v{z}inskis,
S.~Balakauskas, S. A\v{s}montas, and G.~Valu\v{s}is$^{\ddag}$
\\ {\it Semiconductor Physics Institute, A.~Go\v{s}tauto Street 11,
LT-01108 Vilnius, Lithuania}\\
\vspace{0.2truecm} W.-H.~Chow$^{\S}$, P.~Steenson, and P.~Harrison
\\ {\it Institute of Microwaves and Photonics, School of Electronic and
Electrical Engineering, University of Leeds,
Leeds LS2 9JT, United Kingdom}\\
\vspace{0.2truecm} A.~Lisauskas and H.~G.~Roskos
\\ {\it Physikalisches Institut, Johann Wolfgang Goethe-Universit\"{a}t,
Max-von-Laue-Strasse 1, D-60438 Frankfurt/M, Germany}\\
\vspace{0.2truecm} K.~K\"{o}hler
\\ {\it Fraunhofer-Institut f\"{u}r Angewandte Festk\"{o}rperphysik,
Tullastrasse 72, D-79108 Freiburg, Germany}}

\vskip 1.0 truecm


\begin{abstract}

Electric field gradient effects induced by an asymmetrically in-plane
shaped GaAs/AlGaAs modulation-doped structures of various design
are investigated within 4--300~K temperature range. It is demonstrated 
that current--voltage characteristics of such structures at low,
4--80~K, temperatures exhibit well-pronounced asymmetry arising due
to a presence of two different gradients of the electric
field in a two dimensional electron gas. This phenomenon is caused by both, different accumulation of two-dimensional electrons due to asymmetrical shape of the structure and nonlocality in the electron drift velocity. Experiments are illustrated by a phenomenological model and Monte Carlo simulation. Possible applications of the effect to detect electromagnetic radiation of GHz and THz frequencies are discussed as well.

\end{abstract}
\maketitle


\section{\label{sec:level1}Introduction}

As it is known, in modulation-doped (or selectively-doped)
structures free electrons, spatially separated from ionized
donors, are confined to a narrow potential well forming the so-called
two-dimensional electron gas (2DEG)\cite{stormer1979,
hiyamizu1983}. $Along$ the $z$ axis, i.~e. perpendicular to the
2DEG plane, the confinement potential has a triangular shape, and
the spectrum of the 2DEG is composed of a sequence of quantum
subbands.\cite{ando1982} The shape of the confining potential has
influence only to the carrier transitions between different
subbands; meanwhile for excitations within each subband --
so-called intraband excitations -- this effect is very small.
$Perpendicular$ to the $z$ axis, i. e. in the 2DEG, or $x-y$
plane, the carriers display very high electron mobilities exceeding
values of 2$\times10^{5}~\rm{cm}^2$/(V$\cdot$s) at liquid nitrogen
and 1$\times10^{6}~\rm{cm}^2$/(V$\cdot$s) at liquid helium
temperatures.\cite{heiblum1984} This fundamental property of these
structures allowed to invent HEMTs -- high electron mobility
transistors.\cite{mimura1980}

The carrier transport in 2DEG layers at low electric fields is described by the Ohm's law together with the Einstein relation for the diffusion coefficient, i.e.
\begin{eqnarray}
v = \mu[E- (kT/e)(1/n)\nabla n], 
\end{eqnarray}

where $v$ is drift velocity, $\mu$ stands for electron mobility, $E$ is the electric field and $n$ is the free carrier concentration; $e$ is the electron charge, $k$ and $T$ denote Boltzmann constant and lattice temperature, respectively. This so-called "drift-diffusion equation" describes the quasi-classical situation when the quantization of the electronic states-induced effects are not important. On the other hand, it does not take into account the hot-electrons phenomena when the applied in-plane electric field is strong enough to heat the electrons up to the energy exceeding their equilibrium value.\cite{price1988}  

The electric field in modelling, as a rule, is assumed to be uniform. Although this assumption is very convienent for calculations, however, in many cases, in order to obtain the real picture of physical processes, it is necessary to include spatial variation of the carrier concentration or/and  the electric field along the channel. In particular, when it becomes comparable with the range of a mean free path. Under these circumtances, effects such as velocity overshoot begin to predominate defining thus the performance of the device. Such an approach is very important, for instance, in understanding the behavior of devices containing asymmetric channel profile produced by a relevant doping. More specifically, in a quarter-micron $n$-type silicon metal-oxide semiconductor field effect transistors (MOSFET) with asymmetric channel profile, formed by the tilt-angle ion-implantation after gate electron formation, allows one to achieve high current drivability and hot-electron reliability.\cite{buti1991} For example, in 0.1~$\mu$m gate length asymmetric $n$-MOSFET structures this technological innovation allows to attain higher electron velocity in comparison with conventional devices of such  type.\cite{odanaka1997} 

In this article, we report on experimental and theoretical investigation of the {\it electric field gradient-induced effects} due to asymmetrically in-plane shaped GaAs/AlGaAs modulation-doped structures. We show that current--voltage characteristics of such structures at low, 4--80~K, temperatures exhibit pronounced asymmetry. The physics behind is attributed to a two-dimensional bigradient effect which is equivalent to the phenomenon observed earlier in a bulk asymmetricaly-shaped semiconductors\cite{asmontas1975, stepas1975}.  We demonstrate that depending on the values of the in-plane electric fields and their gradients, the effect can reveal itself as a result of different distribution of accumulating two-dimensional electrons due to the asymmetrical shape of the structure, and/or the exhibition of nonlocal drift velocity which becomes pronounced in a different manner due to the presence of two different gradients of the in-plane electric field. 

The paper is organized as follows. In Sec.~II we present the
design of the GaAs/AlGaAs modulation-doped structures, their
electrical parameters and geometry features;  we also describe
briefly used measurement techniques. Section~III reports on experimental  
results obtained in GaAs/AlGaAs modulation-doped structures of various designs and at different lattice temperatures. Section~IV is devoted to theoretical models and illustrates the concept of the electric field gradients-induced phenomenon -- the bigradient effect with special emphasis on manifestation of the electron drift velocity. In Sec.~V possible applications of the effect for the sensing of electromagnetic radiation within GHz--THz frequencies are disccused, while in Sec.~VI features of three-dimensional vs. two-dimentional effect are compared. Finally, conclusions are given in Sec.~VII.

\section{\label{sec:level1}Samples and measurement techniques}

Two types of modulation-doped structures of GaAs/Al$_{0.25}$Ga$_{0.75}$As
(structure 2DEG-A) and GaAs/Al$_{0.3}$Ga$_{0.7}$As (structure 2DEG-B) were
grown by molecular beam epitaxy technique. Their design parameters are given in the caption of Fig.~1.

\begin{figure}[t]
\centering\includegraphics[width=7cm,height=5cm]{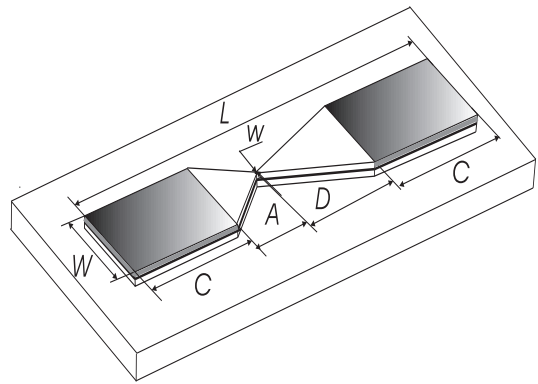}
\caption{ The shape of the studied structures placed on
semi-insulating substrate. Characteristic dimensions of the
structure are the following: $L$=500~$\mu$m; $D$=250~$\mu$m,
$A$=50~$\mu$m, $W$=100~$\mu$m. White color denotes active part
containing 2DEG which is shown schematically as a black sheet; grey
colored parts depict Ohmic contact areas of length
$C$=100~$\mu$m. Layer sequence of GaAs/AlGaAs modulation-doped
structures (from the top): structure 2DEG-A and structure 2DEG-B,
respectively: 20~nm $i$-GaAs cap layer; 80~nm Si-doped, 1$\times
10^{18}~\rm{cm}^{-3}$, layer of Al$_{0.25}$Ga$_{0.75}$As and 60~nm
Si-doped, 2$\times 10^{18}~\rm{cm}^{-3}$ of
Al$_{0.3}$Ga$_{0.7}$As; undoped spacers, 45~nm
Al$_{0.25}$Ga$_{0.75}$As and 10 nm Al$_{0.3}$Ga$_{0.7}$As; 1000~nm
and 600~nm of $i$-GaAs; smoothing superlattice -- twenty and six
periods of 9~nm AlGaAs/1.5~nm-GaAs layers; 0.5~$\mu$m and
0.6~$\mu$m layer of $i$-GaAs; semi-insulating substrate.}
\label{f:1}
\end{figure}

Electron sheet density,  $n_{\rm 2D}$, and low-field mobility,
$\mu$, at different, 300~K, 77~K and 4.2~K, temperatures for
structure 2DEG-A are the following: 5.5$\times 10^{11}~\rm{cm}^{-2}$
and $4700~\rm{cm}^2$/(V$\cdot$s), 1.9$\times10^{11}~\rm{cm}^{-2}$
and $190600~\rm{cm}^2$/(V$\cdot$s), 1.9$\times
10^{11}~\rm{cm}^{-2}$ and 2$\times 10^{6}~\rm{cm}^2$/(V$\cdot$s),
respectively; while for structure 2DEG-B:
5.6$\times10^{11}~\rm{cm}^{-2}$ and $8000~\rm{cm}^2$/(V$\cdot$s),
4.3$\times 10^{11}~\rm{cm}^{-2}$ and
$116000~\rm{cm}^2$/(V$\cdot$s), 4.3$\times 10^{11}~\rm{cm}^{-2}$
and 1$\times 10^{6}~\rm{cm}^2$/(V$\cdot$s), respectively.

The wafers were then processed into asymmetrically-shaped samples
of different geometry which was changed varying the width $w$ of the
neck. We have used the optical lithography for the samples with 
$w$ = 12~$\mu$m.  The meza with height, $d$, of 2~$\mu$m in these samples was fabricated by wet etching. In order to fabricate samples with a narrower neck, respectively 5~$\mu$m, 2~$\mu$m, and 1~$\mu$m width, we have employed electron-beam lithography and shallow wet etching; the meza height in this case was 300~nm. The Ohmic contacts were produced by a rapid annealing  procedure of evaporated Au/Ge/Ni compound.

The wired samples were then mounted into a two-stage closed-cycle helium cryostat, and $I-V$ characteristics were measured within 4--300~K temperature range. In microwave experiments performed at 10~GHz frequency, we have used magnetrons delivering pulses  of 1.5--5~$\mu$s duration with the repetition rate of 35--40~Hz; the samples were then placed into
rectangular waveguides, and the microwave radiation-induced voltage
arising over the ends of the sample was measured by an oscilloscope;
the experiments were performed at room and liquid nitrogen
temperatures. In the terahertz (THz) range, 0.584--2.52~THz, the source of THz radiation was an optically-pumped molecular laser operating in the continuous wave regime; the illumination into the sample, located in the closed-cycle helium cryostat behind the THz transparent high-pressure polythene window, was focused by spherical mirrors. 
As in the microwave case, the external radiation-induced voltage arising over the ends of the sample was recorded by lock-in amplifier at
a chopper frequency of 187~Hz. 

\section{\label{sec:level1}Experimental Results}

Figure~2 shows the experimental results of  $I-V$ characteristics measured in structure 2DEG-A in the samples with different neck widths recorded at liquid helium temperature. (The experimental data of the structure 2DEG-B are equivalent and therefore are not presented here). 

\begin{figure}[t]
\centering\includegraphics[height=6.9cm]{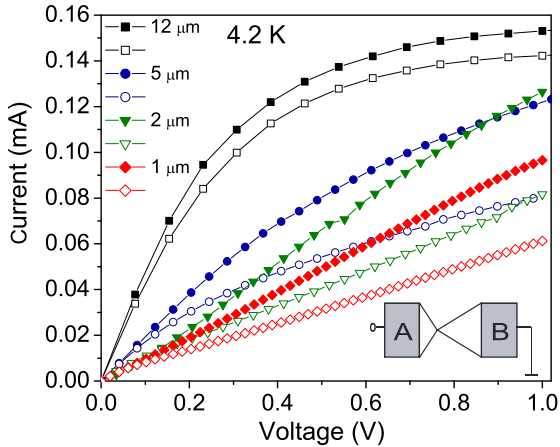}
\caption{(Color online)
 $I-V$ characteristics of the asymmetrically shaped samples with
 different neck width at liquid helium temperatures.
 The inset depicts the measurement scheme:
 the contact~B was always grounded, positive (filled symbols)
 and negative (empty symbols) voltage was applied to the contact~A.}
\end{figure}

As one can see, the salient feature of all the measured  $I-V$ characteristics is an asymmetry becoming well-procounced above 0.1~V. Moreover, the character of the characteristics strongly distingtive to the reduction of the neck width: For the sample with neck width of 12~$\mu$m two parts in the curve can be distinguished -- the initial rising part (below 0.3~V) which bends with voltage and above 0.7~V transits to the second, saturated one. With the decrease of the width $w$ the saturated part transforms to a boosting region with highly expressed asymmetry. A complex behaviour is observed in the 5~$\mu$m width sample:  careful look can resolve initial rising part below 0.15~V, rudiments of saturated part within 0.15--0.6~V and already emerging boosting region above 0.6~V with spread influence on all the curve.  

It should be pointed out that the asymmetry is strongly sensitive to the temperature variation. This is illustrated by the experimental data for 2~$\mu$m neck sample given in Fig.~3. It is seen that at low temperatures the asymmetry is rather large -- if the coefficient of asymmetry of the $I-V$ characteristics is defined as $(I_{f}-I_{r})/I_{f}$ -- at liquid helium temperature under 0.8~V it is 0.506, at 80~K its value is about 0.369, meanwhile at room temperature the asymmetry is nearly invisible. To resolve the effect, the experimental data should be multiplied by a factor of one hundred.  

\begin{figure}[t]
\centering\includegraphics[height=6.9cm]{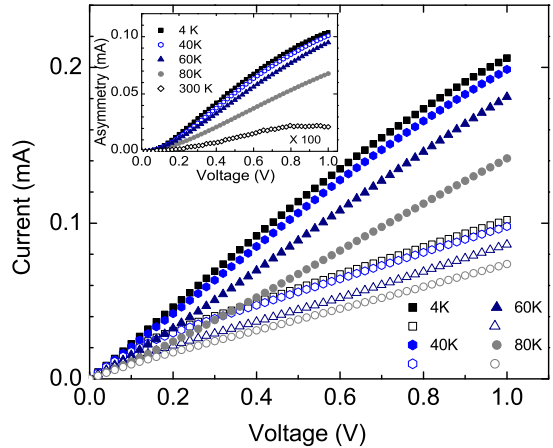} \caption{(Color online) $I-V$ characteristics of the asymmetrically shaped sample vs. temperature.  The neck width is 2~$\mu$m.  Inset shows the change of asymmetry varying the lattice temperature. Measurement scheme is the same as in the previous plot.} 
\end{figure}

\section{\label{sec:level1} Theoretical approach and physics behind}

\subsection{\label{sec:level2}Electric-field gradient-induced effects:
carrier accumulation}

To understand the physics behind the observed phenomenon, we have followed the same ideology as in the bulk case of asymmetrically-shaped structures: We explore an asymmetrically shaped two-dimensional structure by a solving a coupled system of one-dimensional Poisson and current-flow equations neglecting quantum effects\cite{asmontas1975, stepas1975}:

\begin{eqnarray}
{\frac{1}{S(x)}\frac{d}{dx}[S(x)E(x)]=-
\frac{|e|}{\epsilon}\left(n(x)-N_{d}\right)},\\
 I=\{|e|n(x)\mu(E)E(x)+|e|D(E(x))\frac{d}{dx}[n(x)]\}S(x),
\end{eqnarray}

 where $x$ denotes the position along the symmetry axis of
the asymmetrically-shaped sample, $S(x)$ is the cross-section area of the
sample, $n(x)$ is the electron concentration, $N_{d}$ is the donor
concentration (used here as a free fit parameter), $E(x)$ denotes
the electric field strength, $\epsilon$ is the dielectric constant of the sample, $D(E)$ denotes the diffusion coefficient, and $\mu(E)$ is the electron mobility. The diffusion coefficient may be set to a constant value as its dependence on the electric field has no appreciable
influence on the results. As it is known, mobility $\mu(E)$ 
changes with the electric field:\cite{shah1984}
$\mu(E)=\mu_{0}E_{c}(\mu_{0})/[E(x)+E_{c}(\mu_{0})]$, where $E_{c}
= v_{c}/\mu_{0}$, and $v_{c}$ is the cut-off velocity equal to
$10^{7}$ cm/s.

For a given value of the current $I$, one can
employ an iterative scheme to solve self-consistently the
system and calculate the voltage.

Results of the electric field and relative carrier density calculations are given in Fig.~4. The model calculations do give two different gradients of the electric field induced by the asymmetrical shape of the structure depicted in Fig.~1. One must note further that the distribution depends on the current direction, and this difference is appreciable only within several microns from the neck: The electric field peaks are sharper in the case of the reverse current, when the A-contact of the sample is under a positive bias. 

\begin{figure}[t]
\centering\includegraphics[height=6.8cm]{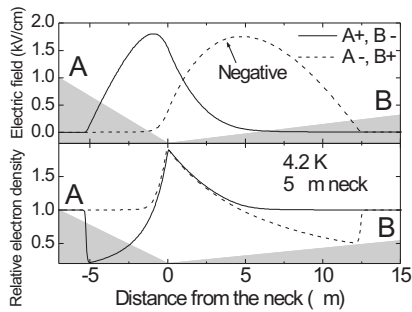}
\caption{(Color online) Calculated electric field and carrier density in the sample with 5~$\mu$m neck width under $I$ = 0.03~mA at liquid helium temperature. }
\end{figure}

Hence, the in-plane geometrical asymmetry induces two different gradients of the electric field in the sample. As a consequence, different spatial accumulation of two-dimensional electrons in the vicinity of the neck can be a reason for asymmetric $I-V$  curves. This prediction is confirmed by the theoretically calculated $I-V$ characteristics and their comparison with the experimental data presented in Fig.~5.  As one can see, the measured and calculated 
$I-V$ curves in 12~$\mu$m neck width samples behave similarly: 
in low electric fields, where Ohm's law is still valid,
the current values are the same and do not depend on the polarity
of the applied voltage. However, with the increase of voltage --
when sample is tuned to a strong field regime -- the electrons become hot, they
accumulate differently in the vicinity of the neck due the
asymmetrical shape of the structure. As a result, the asymmetry in the $I-V$ characteristics is observed. Since this phenomenon relies on hot-carriers, it is reasonable that is temperature-sensitive: the decrease of lattice temperature will make the effect much more
pronounced since the carrier heating at low temperatures is stronger. This
is clearly seen in Fig.~3: the difference between the forward and
the reverse-bias curves, for instance, at room temperature\cite{room} is
approximately by three orders of magnitude smaller than that at
4.2~K temperature in the given voltage range. The temperature effect is also confirmed by theoretical calculations (data are not given here). 

Therefore, the presence of two different electric field gradients in the hot two dimensional electron regime, allows one to explain the asymmetry of $I-V$ characteristics at low temperatures. The phenomenon can be called thus the two-dimensional bigradient effect.

It is reasonable that the reduction of the neck should cause stronger asymmetry in  $I-V$ curves. Experimentally it is clearly illustrated in Fig.~2. However, if one compares the behavior of the curves, for instance, for the 5~$\mu$m neck  sample with a corresponding theoretical estimates (Fig.~5),   
evident differences can be indentified:
\begin{figure}[h]
\centering\includegraphics[height=9.1cm]{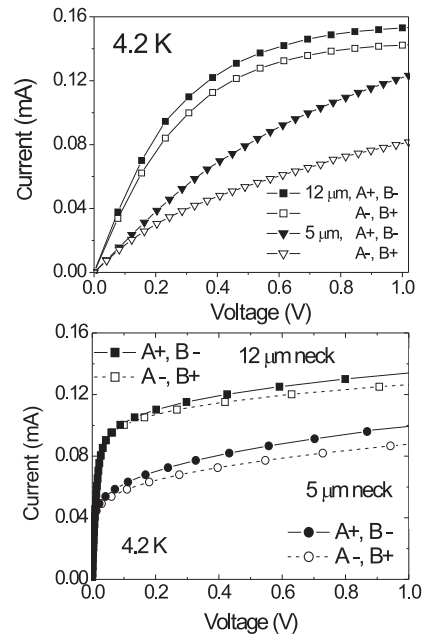}
\caption{Comparison of the measured (top plot) and calculated $I-V$ (bottom plot) characteristics in the asymmetrically-shaped samples with different -- 12~$\mu$m and  5~$\mu$m -- neck widths at liquid helium temperature.}
\end{figure}

the theory does not follow the experiment anymore neither in curve shape nor in asymmetry value. Therefore, in order to understand the origin of the behavior, a more complete theoretical model must be applied.  

\subsection{\label{sec:level2} Electric-field gradient-induced effects: 
Influence of nonlocality in drift velocity}

Althought the equation (1) is the backbone of electronic device simulations, in the current situation one needs to be extended taking into account the electric field gradient-induced term.\cite{price1988}  The electron average velocity can be then expressed using the extended/augmented drift-diffusion model\cite{sonoda1996}    

\begin{eqnarray}
v=v_{h}(1+\frac{\delta}{E}\frac{dE}{dx}) - \frac{D}{n}\frac{dn}{dx},
\end{eqnarray}

where $v_{h}$ is the drift velocity in a homogeneous electric field, $D$ denotes diffusion coefficient, $\delta$ depics the length coefficient. The latter parameter is in the order of electron mean free path and determines the severity of the velocity overshoot effect which is pronounced when the variation range of the electric field is equal to or shorter than $\delta$. Also, it is worth noting that in the given equation we have assumed that nonlocality in diffusion coefficient is negligible and has no observable influence on the effect.  

A point of departure for analyzis is an estimate of a mean free path ($\lambda$) of two dimensional electrons at liquid helium temperatures. As electron mobility in the studied structures is of about 1$\times 10^{6}~\rm{cm}^2$/(V$\cdot$s),  this results to the value of 

\begin{eqnarray}
\lambda = \frac{\mu}{e} (\sqrt{3kTm^{*}}) \approx 1~\mu m, 
\end{eqnarray}

where $m^{*}$=0.068$m_{e}$, the latter is the free electron mass. Meanings
of other symbols are the same as in previous formulae.

The second step in considering the data is an estimate of the distribution 
of the electric field in the vicinity of the neck. Voltage value of 0.25~V 
was chosen as a reference point, since experimentally observed asymmetry of the $I-V$ characteristic starts to be strongly expressed. Theoretical calculations using the set of equations (2 -- 3) are given in Fig.~6.  As one can see, the shape of the
\begin{figure}[h]
\centering\includegraphics[height=6.2cm]{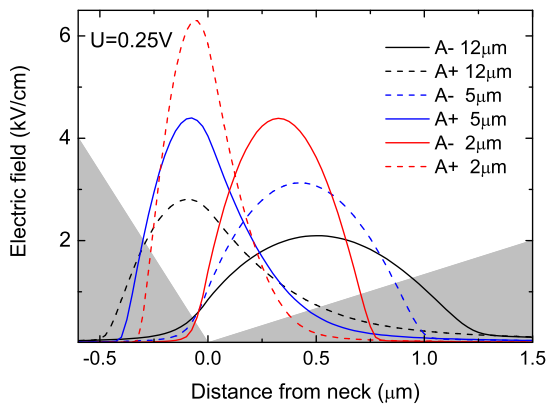}
\caption{(Color online) The dependence of the calculated electric field
distribution in the vicinity of the asymmetrically-shaped sample's
neck for two opposite currents at 4.2~K. The background is shaded in the
form of the sample. Note the drastically increased asymmetry in the
electric field distribution with the reduction of the neck width.}
\end{figure}

    electric field distribution strongly varies with the reduction of the neck width: it becomes sharper and the asymmetry becomes more pronounced. Futhermore, distribution of the electric field is squeezed to the narrower range. If for 12~$\mu$m width sample the electric field extends over 1.7~$\mu$m, in the 5~$\mu$m  and 2~$\mu$m necks it is of about 1.2~$\mu$m and 0.9~$\mu$m, respectively. One can assume therefore that in 5~$\mu$m neck samples the nonlocality can start to predominate since the change of the electric field along the structure is comparable with the electron balistic length. 

Calculations illustrating the effect of nonlocality in electron drift velocity on $I-V$ characteristics are given

\begin{figure}[h]
\centering\includegraphics[width=6.4cm,height=8.4cm]{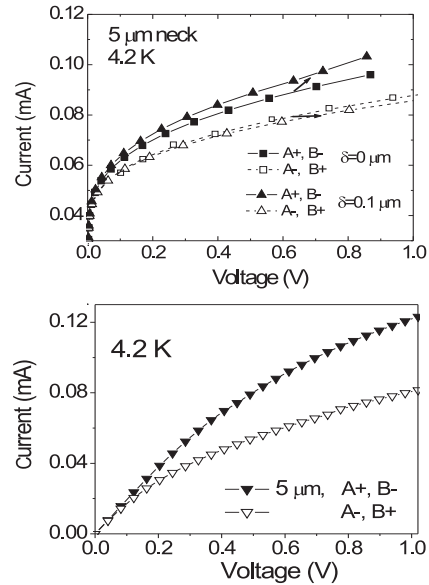}
\caption{Top plot: Calculated $I-V$-characteristics of 5~$\mu$m neck sample using different length coefficient  $\delta$. Note the increase of the asymmetry with  $\delta$ in the dependences.  Bottom plot: Experimental results in 5~$\mu$m neck sample given for comparison.}
\end{figure}

in Fig.~7. One can see that increase of  $\delta$ has obvious effect in 5~$\mu$m neck sample -- even small increase in length coefficient (in the range of  0.1--0.2~$\mu$m) stimulates the raise in the asymmetry. It is evident that calculations for the 5~$\mu$m are very similar to the given experimental results. It is deserve to remark that in 12~$\mu$m neck samples such the influence is not observed (these theoretical data are omitted here).

The inherent feature of the nonlocal effects is a manifestation of the velocity overshoot. To evidence the latter we have performed the Monte Carlo calculations of the drift velocity in a "frozen" electric field taken from a solution of the coupled set of Poisson and current-flow equations (Eq.~2-3) as  described previously. We have simulated the drift of Monte Carlo particles along the x-direction in the given electric field distribution assuming that the particles enter the modelling region having the Boltzmann distribution at the lattice temperature. The standard three-valley GaAs model is applied in the simulation\cite{hess-1984} with an impurity scattering rate corresponding to a $10^{16}~\rm cm^{-3}$ donor concentration. Figure~8 presents simulated electron velocity profiles in forward and reverse directions at the electric field profiles for 5~$\mu$m neck samples. For the sake of evidence, the electric field profiles taken from Fig.~6 are given, too. As one can see, the velocity overshoot/undershoot and strong nonlocality effects are clearly visible. 

\begin{figure}[t]
\centering\includegraphics[width=7.6cm,height=6.6cm]{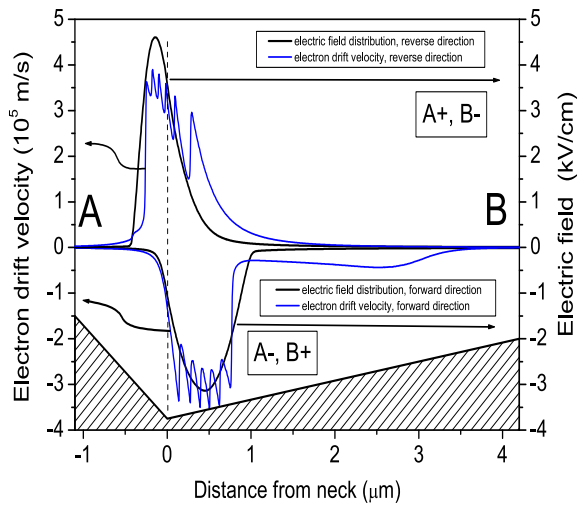}
\caption{(Color online) Monte Carlo calculations of the electron drift velocity in "frozen" electric field in 5~$\mu$m neck sample. Note oscilations in the drift velocity caused by optical phonon emission. The vertical dashed line indicates the neck position. Background is decorated schematically as the geometrical shape of the structure.}
\end{figure}

\begin{figure}[h]
\centering\includegraphics[width=7.1cm,height=8.1cm]{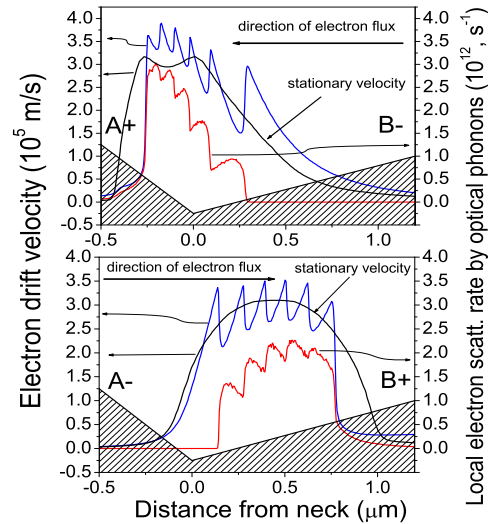}
\caption{(Color online) Monte Carlo calculations of drift velocity and local scattering rate of electrons by optical phonons in "frozen" electric field in 5~$\mu$m neck sample. Note opposite phases in the velocity and the scattering rate. Background schematically shows the geometrical shape of the structure.}
\end{figure}

The sharp peaks at the top of the velocity profiles are related to the locality in optical phonon emission rate $\nu_{op}$. Note that the velocity profile is asymmetric due to the presence of the different electric field gradients and exhibits nearly constant value in the forward direction ranging from 0.75 up to 3~$\mu$m. The Monte Carlo simulation for electric field profiles from Fig.~8 gives the coefficient of current asymmetry of 0.48. The factor strongly depends on impurity scattering rate: it decreases as the scatering rate increases. The simulation with donor concentrations 2$\times 10^{15}$, 1$\times 10^{16}$, and 5$\times 10^{16}~\rm cm^{-3}$ gives the coefficient values of 0.62, 0.48, and 0.25, respectively.

To get close-up view in the physics behind, we have depicted the velocity and $\nu_{op}$ profiles in both, forward and reverse, directions (Fig.~9).
Also, stationary velocity is added to illustrate the overshoot/undershoot areas. It is obvious that the electron velocity minima coincide with the $\nu_{op}$  maxima forming thus "comb" in electron distribution. The effect is equivalent to the the free-carrier grating formation phenomenon caused by locality of optical phonon emission as already demonstrated in $n^{+}nn^{+}$ InN structures.\cite{viktor-2004}

\section{\label{sec:level1}possible applications of the effect}

Since the two-dimensional bigradient effect induces asymmetry in
$I-V$-characteristics, it is reasonable to consider its possible
applications in detection of the high-frequency electromagnetic
radiation. To test this idea, we have placed samples in a
microwave field of 10~GHz frequency and recorded microwave-induced
voltage arising over the ends of the sample. The results for two
samples of different neck width recorded at room and liquid
nitrogen temperatures are displayed in Fig.~10(a).

\begin{figure}[h]
\centering\includegraphics[width=7.7cm,height=5.9cm]{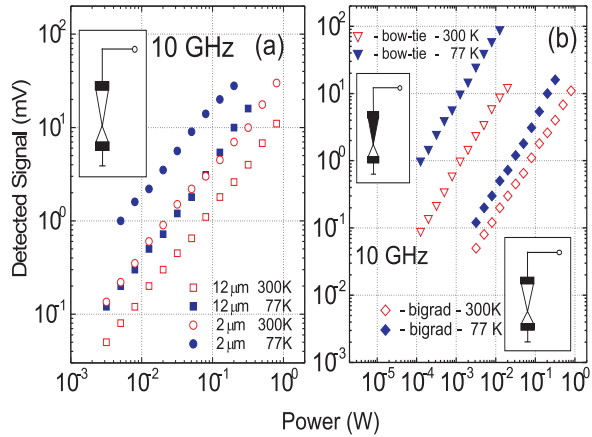}
\caption{(Color online) Detected signal versus incident power at
10~GHz frequency in the samples fabricated from the structure 2DEG-B.
(a) -- Data for two samples employing two-dimensional
bigradient effect with neck sizes of 12~$\mu$m and 2~$\mu$m at
room and liquid nitrogen temperatures. (b) -- Comparison of
the two-dimensional bigradient effect-based samples with bow-ties
diodes containing 2DEG layer. Geometrical dimensions and are the
same for both type of samples. Data recorded in 12~$\mu$m neck
size samples at room and liquid nitrogen temperatures. Insets show
measurement schemes and design differences. Note much higher
sensitivity of the 2DEG bow-ties diodes.}
\end{figure}

One can see that with the increase of incident power the signal
increases linearly in all the range of studied power. Thus, due to
the asymmetry in $I-V$ curves caused by the two-dimensional
bigradient effect the microwave radiation can be detected
\cite{THz-detect}. However, as estimates show, the sensitivity at
room temperature is rather low. For instance, it is only about
12~mV/W and 40~mV/W for the samples with neck width of 12~$\mu$m
and 2~$\mu$m, respectively. Since the effect is related to the
electron heating, the decease in temperature should induce the
increase in the detected signal. The temperature effect, however,
is not as large as might be expected from hot-electron physics
where change of the signal is proportional to the carrier
mobility. As one can see, at 77~K, the detected voltage increases
reaching the values of 40~mV/W (12~$\mu$m neck sample) and
150~mV/W (2~$\mu$m neck sample). In fact, the rise in only by a
factor of about 4, while the mobility increases about 14
times. We associate the outcome with a presence of ballistic
electrons which cross the effective length of the samples
without scattering.\cite{shur2002} It is evident that for
implementation of the effect into practical devices, such a
sensitivity values are not sufficient. Following the ideology of
the bigradient effect, one needs to produce much higher
non-uniformity of the electric field in the sample's neck
vicinity. This can be achieved easily by metallizing one of the asymmetric
parts of the structure. This is evidenced by calculations of the
distribution of the electric field given in Fig.~11.

\begin{figure}[h]
\centering\includegraphics[height=6.7cm]{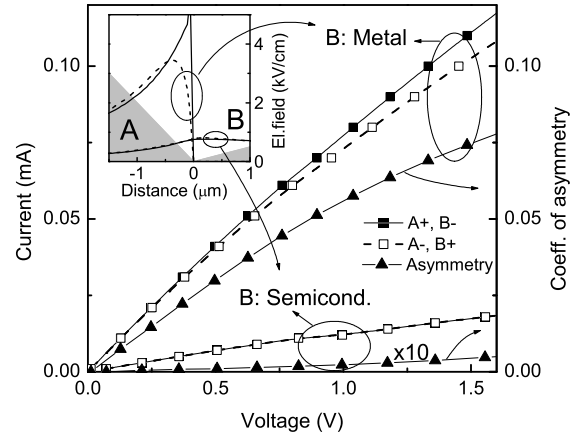}
\caption{Calculated effect of the metallization of the B-part of
the asymmetrically shaped sample on $I-V$ characteristics.
Room temperature value of the electron mobility $\mu$=
$4700~\rm{cm}^2$/(V$\cdot$s) is used. The neck width is 5 $\mu$m.
The filled squares denote the forward current $I_{f}$, the empty
squares - the reverse current $I_{r}$. The coefficient of
asymmetry is defined as $(I_{f}-I_{r})/I_{f}$.}
\end{figure}

It is seen that the non-uniformity of the electric field with one part
of the sample metallized is much stronger than in the
conventional asymmetrically-shaped structure. It is therefore
reasonable to expect significantly higher values of the
sensitivity. The experimental illustration is given in Fig.~10(b)
where the bigradient structures are compared with the so-called 2DEG
bow-ties diodes\cite{bow-ties}-- structures where one of the
parts of the sample is covered by metal. One can see that these
devices are nearly in two orders of magnitude more sensitive
compared to the structures of bigradient-design.

Of particular interest for the direct applications are the values
of the voltage sensitivity at room temperature. In order to increase
the sensitivity of the diode, we have chosen the design 2DEG-B
(due to the higher value of the sensitivity at room temperature)
and reduced the size of the apex down to 2~$\mu$m keeping the
other geometrical dimensions the same (Fig.~12). As one can see,
in this case the detected signals are significantly larger
than for the 12~$\mu$m neck diodes at 300~K, and the sensitivity
amounts to 2.5~V/W, which is nearly one order of magnitude higher
compared to the diodes of 12~$\mu$m neck width, where it is
about 0.32~V/W. The decrease in temperature down to 77~K allows
one to increase the sensitivity up to nearly 40~V/W, i.e. close to
a factor of 14.5 which corresponds to the change of 2DEG mobility.
The latter effect -- via thickness of spacer -- is also nicely
expressed in Fig.~12 were the detected signal in both devices,
fabricated from 2DEG-A- and 2DEG-B-type structures is compared.

Additionally, it deserves mentioning that the further decrease of the neck
size down to 1~$\mu$m -- 800~nm, gives, however, no increase in
the sensitivity (the data are not presented here). We attribute it
to weaker coupling of the microwave field into the structure with
the reduction of the neck width.

\begin{figure}[h]
\centering\includegraphics[height=7cm]{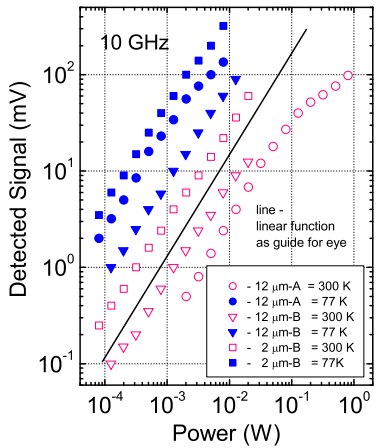}
\caption{(Color online) Detected signal as a function of incident
power at 10~GHz frequency microwave field in 2DEG bow-ties diodes
of different geometry and design. Data recorded in 12~$\mu$m and
2~$\mu$m neck size samples based on structures 2DEG-A and 2DEG-B at room and liquid nitrogen temperatures. Linear function is also given as a
guide for an eye.}
\end{figure}

As to dynamical range, the devices can be used to detect rather different power levels. As already indicated above, the bigradient modulation-doped structures at 300~K in microwaves exhibit sensitivity of about 12~mV/W and 40~mV/W for the samples with neck width of 12~$\mu$m
and 2~$\mu$m, respectively, however, the sensing range is below 1~W in a pulse mode, i.e. the resistance to overloading is small. Bulk silicon-based 12~$\mu$m neck width structures are also not sensitive -- only about 0.1~V/W -- but they can withstand up to 10~kW maximum applied power at room temprature without any negative outcome.\cite{algis1999} Asymmetrically-shaped bulk GaAs structures containing $n-n^{+}$-junction were found to be suitable for broadband sensing within GHz--THz frequencies at 300~K.\cite{algis-gintaras-jap2003}

Finally, it is worth to remark on the operation of the bigradient structures
within the THz range. One can note that at 300~K the sensitivity
at 0.763~THz in the sample with neck width of 12~$\mu$m it is
about 13~mV/W, i.e. close to the value in 10~GHz range. Narrowing
the neck width down to 2~$\mu$m, the sensitivity at 0.763~THz can
increase up to 17~mV/W. Within frequencies 1.41~THz, 1.63~THz and
2.52~THz the detection observed is rather poor. We explain it by a
weak coupling of the incident THz field into the structure. At
helium temperatures, the sensitivity reaches the value of 60~mV/W
at 0.763~THz for the 12~$\mu$m neck width sample.\cite{estimate}

\section{\label{sec:level2}Two-dimensional vs. bulk bigradient effect}

The effect was originally discovered by measuring $I-V$ characteristics in
bulk semiconductors, $n$-Ge and  $n$-Si, at room and liquid
nitrogen temperatures.\cite{asmontas1975, stepas1975} 
In what follows we compare the two-dimensional bigradient effect observed in asymmetrically in-plane shaped GaAs/AlGaAs modulation-doped structures with its bulk equivalent desribed in details.\cite{asmontas1985}

\begin{figure}[h]
\centering\includegraphics[height=6.1cm]{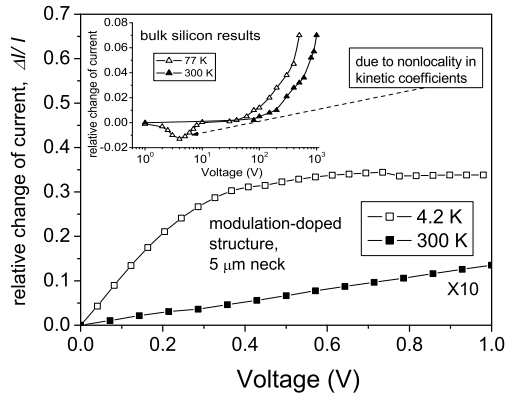}
\caption{Change of the relative current in 5~$\mu$m neck size sample at liquid helium and room temperatures. Insert depicts the same plot for comparison in $n-$doped asymetrically-shaped silicon samples with specific resistance of 2.5~$\Omega\cdot$cm at 300~K. The length of the sample is 23~mm, angles of the asymmetry are $\alpha_{1}$=$45^0$ and $\alpha_{2}$=$13^0$. More details can be found in Ref.~19.} 
\end{figure}

The inherent fearure of both -- the two-dimentional and bulk -- effects is asymmetry in $I-V$ characteristics. Different material parameters and geometrical dimensions of the samples determine rather distinctive voltage scale and temperature conditions for experimental observation. The effect of the dimensionallity we associate with unusual manifestation of the nonlocality in drift velocity. Figure~13 presents a change of the relative current in 5~$\mu$m neck size sample at 4.2~K and 300~K. For comparison of the features, inset shows equivalent bulk data measured in $n$-type silicon according to Ref.~19.   

It is seen that in the bulk semiconductor the nonlocality is associated with the sign change in the asymmetry clearly observed within 1-10~V voltage scale at 77~K (see inset). In contrast, in modulation-doped structure, no such  peculiarity is observed. It is worth noting, however, that above 0.3~V the change of the relative current is nearly independent of voltage -- as was indicated earlier (Fig.~7, above 0.25~V range, where asymmetry of $I-V$ characteristics strongly increases) -- we attribute it to the apperance of the nonlocal drift velocity.

\section{\label{sec:level1}Conclusions}

Asymmetrically in-plane shaped GaAs/AlGaAs modulation-doped
structures of various design are studied within 4--300~K
temperature range. We have shown that current--voltage
characteristics of such structures at low, 4--80~K, temperatures
exhibits pronounced asymmetry which we explain by a
two-dimensional bigradient effect. This phenomenon can be
decomposed into two constituents -- carrier accumulation and nonlocality in the electron drift velocity -- each of which manifests itself
depending on the values of the in-plane electric fields and their
gradients. Varying these parameters via change of the sample neck
we have inferred that the effect is induced by different accumulation of
two-dimensional electrons due to asymmetrical shape of the
structure when the neck size down is 12~$\mu$m; in 5~$\mu$m neck
size structures in electric fields close to 3~kV/cm, the nonlocality of the electron drift velocity becomes  predominating causing thus further increase of the asymmetry in $I-V$ curves.
The theoretical models suggested conform well with the obtained
experimental results. Monte Carlo simulation has shown strong nonlocal effects
in the electron drift velocity and presence of overshoot/undershoot 
regions depending on the electric field and its gradient values.
The features of the two-dimensional bigradient effect and its bulk equivalent are compared and possible applications of this phenomenon to detect electromagnetic radiation of GHz and THz frequencies are discussed as well.

\begin{acknowledgments}
We sincerely grateful to Vincas Tamo\v{s}i\={u}nas, Jurek
Lusakowski and Kirill Alekseev for valuable and comprehensive discussions; 
our highest possible appreciations go to Adolfas Dargys for his
illuminating remarks and delicate suggestions in the manuscript
preparation, and Antanas Reklaitis for useful notes on simulation matters.  

The work was supported, in part, by the NATO SfP project 978030
"Broadband detectors". The research at the Semiconductor Physics
Institute at Vilnius was performed under the topic "Study of
semiconductor nanostructures for terahertz technologies" (No.
144.3).
\end{acknowledgments}



\begin{thebibliography}{99}
\bibitem{room}
It is worth noting that the different-bias curves at room
temperature separate noticeably only at voltages above 10~V.
\bibitem{estimate}
The sensitivity was estimated scaling the spatial profile of the THz beam, calculating power density and multiplying it by area of the structure. It was assumed that all the incident power is absorped by the sample.
\end{thebibliography}
\end{document}